\documentclass[a4paper,12pt]{article}

\usepackage[a4paper,left=80pt,right=80pt,top=90pt,bottom=98pt]{geometry}
\usepackage[latin1]{inputenc}
\usepackage{dsfont}
\usepackage{amsfonts,amsmath,amssymb}
\usepackage{amsthm}
\usepackage{graphicx}
\usepackage[small,bf]{caption}
\usepackage{subcaption}
\usepackage{booktabs}
\usepackage[numbers, sort]{natbib}
\allowdisplaybreaks[1]
\usepackage{color}
\usepackage{ulem}
\usepackage{tikz}
\usepackage{multirow}
\usetikzlibrary{arrows}
\usepackage{xspace}
\def\bal#1\eal{\begin{align}#1\end{align}}
\newcommand{\be}{\begin{equation}}
\newcommand{\ee}{\end{equation}}
\newcommand{\bea}{\begin{eqnarray}}
\newcommand{\eea}{\end{eqnarray}}
\newcommand{\besub}{\begin{subequations}}
\newcommand{\eesub}{\end{subequations}}
\newcommand{\ba}{\begin{array}}
\newcommand{\ea}{\end{array}}
\newcommand{\bi}{\begin{itemize}}
\newcommand{\ei}{\end{itemize}}

\begin{document}

\begin{titlepage}


\vspace*{1cm}

\begin{center}
{\Large 
{\bf
Relativistic Freeze--in
}}
\\
[2cm]

{
{\bf
Oleg Lebedev$^1$ and  Takashi Toma$^2$
 }}

\vspace{1.5cm}
 {\it {\small  
 $^1$Department of Physics, University of Helsinki,\\ Gustaf H\"allstr\"omin katu 2a, Helsinki, Finland

 \vspace{0.3cm}
 $^2$Department of Physics, McGill University,\\
 3600 Rue University, Montr\'{e}al, Qu\'{e}bec H3A 2T8, Canada\\ }}

\end{center}

\vspace*{0.5cm}

\vspace*{0.4cm}

\vspace*{1.2cm}

\begin{abstract}
\noindent
 We study production of scalar dark matter via the freeze--in mechanism in the relativistic regime, focussing on the simplest Higgs portal model. 
 We derive the corresponding relativistic reaction rates based on  the Bose--Einstein statistics taking into account the thermal mass effects as well as the change in the Higgs
 degrees of freedom at the electroweak phase transition.
 The consequent constraints on the Higgs portal coupling are obtained.
      \end{abstract}


\end{titlepage}
\newpage

\tableofcontents

\section{Introduction}

The freeze--in dark matter (DM) production mechanism \cite{McDonald:2001vt},\cite{Hall:2009bx} is an attractive possibility for generating the observed DM abundance.
It makes use of a feeble coupling between the Standard Model (SM) and dark matter,
which leads to slow production of DM by the SM thermal bath. In this case, dark matter never thermalizes and its abundance accumulates over time. 
DM production can take place both in relativistic and non--relativistic regimes, depending on the dominant mode for a given dark matter mass.

The existing analyses typically employ the non--relativistic approximation replacing the Bose--Einstein distribution function with the Maxwell--Boltzmann one.
While  this can often  be justified, it can also significantly underestimate the reaction rates at high temperatures. Depending on the model parameters,
the rates can differ by orders of magnitude  \cite{Arcadi:2019oxh}.
In this work, we derive the relativistic reaction rates  based on the Bose--Einstein statistics in the framework of Higgs portal dark matter \cite{Silveira:1985rk}-\cite{Burgess:2000yq}.
These calculations parallel our recent work on self--interacting scalar dark matter \cite{Arcadi:2019oxh}.
We also take into account the effects of thermal mass corrections and the electroweak phase transition, both of which make a significant  impact on the result.

 Recent work on the Higgs portal dark matter freeze--in which employs various degrees of simplification
 can be found in Refs.\;\cite{Yaguna:2011qn}-\cite{Im:2019iwd}.\footnote{Ref.\;\cite{Belanger:2018mqt} obtains a different parametrization for the relativistic rates. However, it does not take into account the thermal mass corrections nor the effect of the electroweak phase transition.}

\section{The set--up}

We consider the simplest real scalar dark matter model with the Higgs portal interaction
\begin{equation}
 V_{hs} = {1\over 2}\lambda_{hs} H^\dagger H s^2   \;,
 \end{equation}
 where $s$ is a real scalar with the potential
 \begin{equation}
 V_{s} = {1\over 4!} \lambda_{s}  s^4 + {1\over 2 } m_{s0}^2 s^2  \;.
  \end{equation}
 Here, the symmetry $s\rightarrow -s$ stabilizes the scalar which plays the role of dark matter (DM). 
The freeze--in DM production takes place in the Standard Model thermal bath through the Higgs coupling when
 $\lambda_{hs} \ll 1$.  For the coupling values in the ballpark of $\lambda_{hs} \sim 10^{-11}$, one obtains the correct 
 DM relic abundance, while DM never thermalizes (assuming also  that $\lambda_s$ is small enough).

 There are two reactions in which DM is produced: Higgs annihilation  $hh \rightarrow ss$ and Higgs decay $h \rightarrow ss$, if allowed kinematically. The latter is only possible at temperatures below the electroweak scale,
 so we distinguish
 \\ \
 
 (a) symmetric phase, $T \gtrsim T_{\rm EW}$ and $\langle H \rangle =0$
 
 (b) broken phase, $T < T_{\rm EW}$ and $\langle H \rangle \not =0$
 \\ \ \\
 Here $T_{\rm EW} \sim v$ is the EW phase transition temperature. In the symmetric phase, the gauge bosons are massless and there are 4 massive Higgs degrees of freedom  $h_i$ such that
 \begin{equation}
 V_{hs} = {1\over 4}\lambda_{hs}  \sum_i h_i^2 s^2   \;. 
  \end{equation}
  In this case, dark matter is produced through annihilation,
  \begin{equation}
 h_i h_i \rightarrow ss \;.
  \end{equation}
In the broken phase, one can use the unitary gauge $H= (0,(h+v)/\sqrt{2})^T$, which is singular at $\langle H \rangle =0$, such that 
  \begin{equation}
 V_{hs} = 
 {1\over 4} \lambda_{hs} h^2 s^2  +  {1\over 2} \lambda_{hs} v \;h s^2 + {1\over 4} \lambda_{hs} v^2   s^2 \;.
 \end{equation}
  This allows for both annihilation and decay, if allowed kinematically,
  \begin{equation}
 hh \rightarrow ss ~~,~~ h \rightarrow ss \;.
  \end{equation}
Note that only one Higgs degree of freedom contributes in this case.\footnote{The would-be Goldstone boson contributions are recovered in the high energy limit through longitudinal  
gauge boson scattering. In this case, the Higgs propagator is cancelled by the gauge boson  polarization vectors producing an effective quartic vertex.}
 
 The total DM yield is given by the sum of yields in the two regimes. Here we omit complications having to do with the transition period at $T \sim T_{\rm EW}$ and approximate the reaction rates using the step function $\theta (T-T_{\rm EW})$.
While in the broken phase relativistic effects are unimportant, at high temperature  they are significant. In this case, it is important to take into account the thermal mass corrections
which represent the leading thermal effects. For the Higgs field, the corrected mass--squared is 
 \begin{equation}
m_h^2 \simeq m_{h0}^2 +  \left( {3\over 16}g_2^2  + {1\over 16}g_1^2  + {1\over 4} y_t^2  
+ {1\over 2} \lambda_h  \right) T^2 \;,
\end{equation}
with $m_{h0}$ being the zero temperature Higgs mass; $g_{1,2},y_t$ are the gauge and  top quark Yukawa couplings, and $\lambda_h$ is the Higgs self--coupling.
In the symmetric phase, $m_{h0}^2= -\lambda_h v^2$ in the convention $V_h=\lambda_h (h^2-v^2)^2/4$.  
On the other hand, we assume that DM is not thermalized and $m_{s0}$
does not receive significant thermal corrections (which would not be suppressed by $\lambda_{hs}$). 

We also note that the DM mass changes during the phase transition and receives an extra contribution, 
\begin{equation}
   m_s^2 =m_{s0}^2 +{1\over 2} \lambda_{hs} v^2 \;.
\end{equation}
For $\lambda_{hs}$ in the range of interest, this effect is negligible unless $s$ has  an MeV (or below)  mass. However, for such light dark matter only the decay production mode
in the broken phase is important, so only the total $m_s^2$ matters. For heavier DM, we make no distinction between $m_{s0}$ and $m_{s}$.

\section{Relativistic reaction rates}

In this section, we compute the $2 \rightarrow 2 $ and $1\rightarrow 2$ relativistic reaction rates necessary for evaluation of the DM relic abundance.
We follow closely our earlier work \cite{Arcadi:2019oxh} where analogous computations for self--interacting scalar DM have been performed.

The  ${a\rightarrow b}$ reaction rates per  unit volume  are 
\begin{equation}
\Gamma_{a\rightarrow b} = \int \left( \prod_{i\in a} {d^3 {\bf p}_i \over (2 \pi)^3 2E_{i}} f(p_i)\right)~
\left( \prod_{j\in b} {d^3 {\bf p}_j \over (2 \pi)^3 2E_{j}} (1+f(p_j))\right)
\vert {\cal M}_{a\rightarrow b} \vert^2 ~ (2\pi)^4 \delta^4(p_a-p_b) . 
\label{Gamma}
\end{equation}
Here ${\cal M}_{a\rightarrow b}$ is the  QFT transition amplitude,  in which we also absorb the {\it  initial and final} state symmetry factors;  
$f(p)$ is the Bose--Einstein momentum distribution function. In thermal equilibrium, $f(p)$ can be written in a covariant form as
\begin{equation}
f(p)= {1 \over e^{u\cdot p \over T} -1 } \;,
\end{equation} 
where $u_\mu$ is the 4--velocity of our reference frame relative to the gas rest frame in which $u=(1,0,0,0)^T$.  

For freeze--in production of DM, the final state enhancement factors $1+f(p_j)$ can be set to 1 since DM is not thermalized and its abundance is much lower than that in equilibrium.

\subsection{ $hh \rightarrow ss $ reaction}

Consider the reaction $hh \rightarrow ss$ due to the quartic interaction term. Following Gelmini and Gondolo \cite{Gondolo:1990dk}, we  write it as
\begin{eqnarray}
&&\Gamma_{2\rightarrow 2}  =  (2\pi)^{-6} \int d^3 {\bf p_1} d^3 {\bf p_2} ~f(p_1) f(p_2) ~\sigma (p_1,p_2) v_{\rm M\o l}  \;,
 \label{Gamma-sigma}
 \end{eqnarray}
where the M\o ller velocity for the incoming particles  is given by
 \begin{equation}
v_{\rm M \o l}= {F(p_1,p_2) \over E_1 E_2}   \;.
\end{equation}
Here $F(p_1,p_2) = \sqrt{(p_1 \cdot p_2)^2 - m_h^4} $ and the cross section is defined by
\begin{equation}
\sigma (p_1,p_2)= {1\over 4 F(p_1,p_2)} \int \vert {\cal M}_{2\rightarrow 2} \vert^2 (2\pi)^4 \delta^4(p_1+p_2 -k_1 -k_2)
\prod_i {d^3 {\bf k}_i \over (2 \pi)^3 2E_{k_i}}  \;,
\label{sigma-def}
\end{equation} 
where, in our convention, $ \vert {\cal M}_{2\rightarrow 2} \vert^2$ includes  the symmetry factors for identical particles in the final $and$ initial states.

The cross section is conveniently calculated in the center-of-mass  (CM) frame. Following \cite{Arcadi:2019oxh}, we introduce
\begin{equation}
p={p_1+p_2\over 2} ~,~k={p_1-p_2\over 2} ~,
\end{equation}
and use the parametrization in terms of the half--the--CM energy $E$ and rapidity $\eta$,
 \begin{eqnarray}
&& p^0= E \cosh \eta, \nonumber\\ 
&& p^1=E \sinh \eta \sin \theta \sin \phi ,\nonumber\\
&& p^2=E \sinh \eta \sin \theta \cos \phi ,\nonumber\\
&& p^3=E \sinh \eta \cos \theta  ,\nonumber
\end{eqnarray}
where $\theta$ and $\phi$ are the polar and azimuthal angles, respectively. 
This corresponds to $p= \Lambda (E,0,0,0)^T$ with $\Lambda$ being a Lorentz transformation.  
As shown in \cite{Arcadi:2019oxh},
for any $g(p_1,p_2)$,
\begin{equation}
\int {d^3 {\bf p_1}\over 2E_1} {d^3 {\bf p_2}\over2 E_2}  ~g(p_1,p_2)=
2 \int_{m_h}^\infty  dE    ~\sqrt{E^2-{m_h^2 }}  ~E^2 \int_0^\infty  d\eta  ~\sinh^2 \eta  ~ \int d\Omega_p ~d\Omega_k ~g(p_1,p_2) \;,
\end{equation}
where in the integrand one must set $k_0=0, \vert {\bf k}\vert=\sqrt{E^2-m_h^2}$ in $k$-dependent quantities, and $\Omega_{p,k}$ are  the solid angles in $p$- and 
$k$-spaces.
In our case, the angular integrations can be performed explicitly. In the CM frame, the cross section is a function of $E$ only and  the angular dependence appears solely in the   
Bose--Einstein factors,
\begin{eqnarray}
&&u\cdot p_1= E\cosh\eta + \sqrt{E^2 -m_h^2} \sinh\eta  ~\cos\theta_k \;,\nonumber\\
&&u\cdot p_2= E\cosh\eta - \sqrt{E^2 -m_h^2} \sinh\eta  ~\cos\theta_k \;,
\end{eqnarray}
where  $k^3= \vert {\bf k}\vert \cos \theta_k$.  Computing the integrals, we get 
\begin{equation}
\int d\Omega_p  \; d\Omega_k~f(p_1) f(p_2) = 8\pi^2 \; 
{2T\over \sqrt{E^2-m_h^2}  \sinh\eta \;(e^{2E \cosh \eta \over T}-1)} 
\ln {   \sinh { E \cosh \eta +  \sqrt{E^2-m_h^2}  \sinh\eta  \over 2T}  \over 
 \sinh { E \cosh \eta -  \sqrt{E^2-m_h^2}  \sinh\eta   \over 2T}}  
 \;.
\end{equation}

The QFT cross section for the process $hh \rightarrow ss$ is 
\begin{equation}
  \sigma^{\rm CM}(E)=   
{1\over 2!2!} ~ { \lambda_{hs}^2  \over 64 \pi E^2 } ~  {\sqrt{E^2-m_s^2} \over \sqrt{E^2-m_h^2}  }\;,
\end{equation}
where, in our convention, we  include the symmetry factor $1/(2!2!)$  associated with identical particles in the final $and$ initial states.
Inserting this in the reaction rate, we get 
\begin{eqnarray}
  \Gamma_{2\rightarrow 2} &=&  {1\over 2! 2!} \; { \lambda_{hs}^2 T \over 16 \pi^5} \\
&  \times & \int_{m_h}^\infty dE ~E \sqrt{E^2-m_s^2} \int_0^\infty d\eta {    \sinh \eta \over e^{{2E\over T} \cosh\eta }-1}~
\ln {  \sinh    {E\cosh\eta + \sqrt{E^2 -m_h^2} \sinh\eta \over 2T}    \over
\sinh   {E\cosh\eta - \sqrt{E^2 -m_h^2} \sinh\eta \over 2T}  } \;, \nonumber
\end{eqnarray}
where $E $ is   half the center--of--mass energy and we have factored out the symmetry factor $1/(2!2!)$.  
This rate is to be multiplied by 4 in the symmetric phase.
  
\subsection{ $h \rightarrow ss $ reaction}

Let us now consider the decay mode $h \rightarrow ss $.
Again, it is convenient to go over to the rest frame of the decaying particle with momentum $p$, so that
$p= \Lambda (E,0,0,0)^T$. 
Parametrizing  $p$ as we did in the previous subsection, we have
$ {d^3 {\bf p} \over 2E } =  \delta(E^2-m_h^2) \;  \sinh^2 \eta  \;E^3   $ $ \;dE ~d\eta ~d\Omega  $. 
Using  the decay width  $\Gamma$ definition
 \begin{equation}
2m_h \Gamma =  \int  
\left( \prod_{j\in f} {d^3 {\bf p}_j \over (2 \pi)^3 2E_{j}} \right)
\vert {\cal M}_{1\rightarrow 2} \vert^2 ~ (2\pi)^4 \delta^4(p-p_f) \;,
\label{width}
\end{equation}
we find 
\begin{equation}
\Gamma_{1\rightarrow 2} =   { \lambda_{hs}^2 v^2 m_h^2 \over 64 \pi^3   } \sqrt{1-{4m_s^2 \over m_h^2}} 
 \;    \int_1^\infty dx \; { \sqrt{x^2-1} \over e^{{m_h\over T} x} -1}   \;,
\end{equation}
 where we have used 
\begin{equation}
 \Gamma= 
  { \lambda_{hs}^2 v^2 \over 32 \pi m_h   } \sqrt{1-{4m_s^2 \over m_h^2}} \;,
 \end{equation}
which includes a symmetry factor $1/2!$ due to identical particles in the final state. In the non--relativistic Maxwell--Boltzmann limit, this agrees with the result in \cite{Hall:2009bx}.

\subsection{Implications}

\begin{figure}[h!]
\begin{center}
 \includegraphics[scale=0.63]{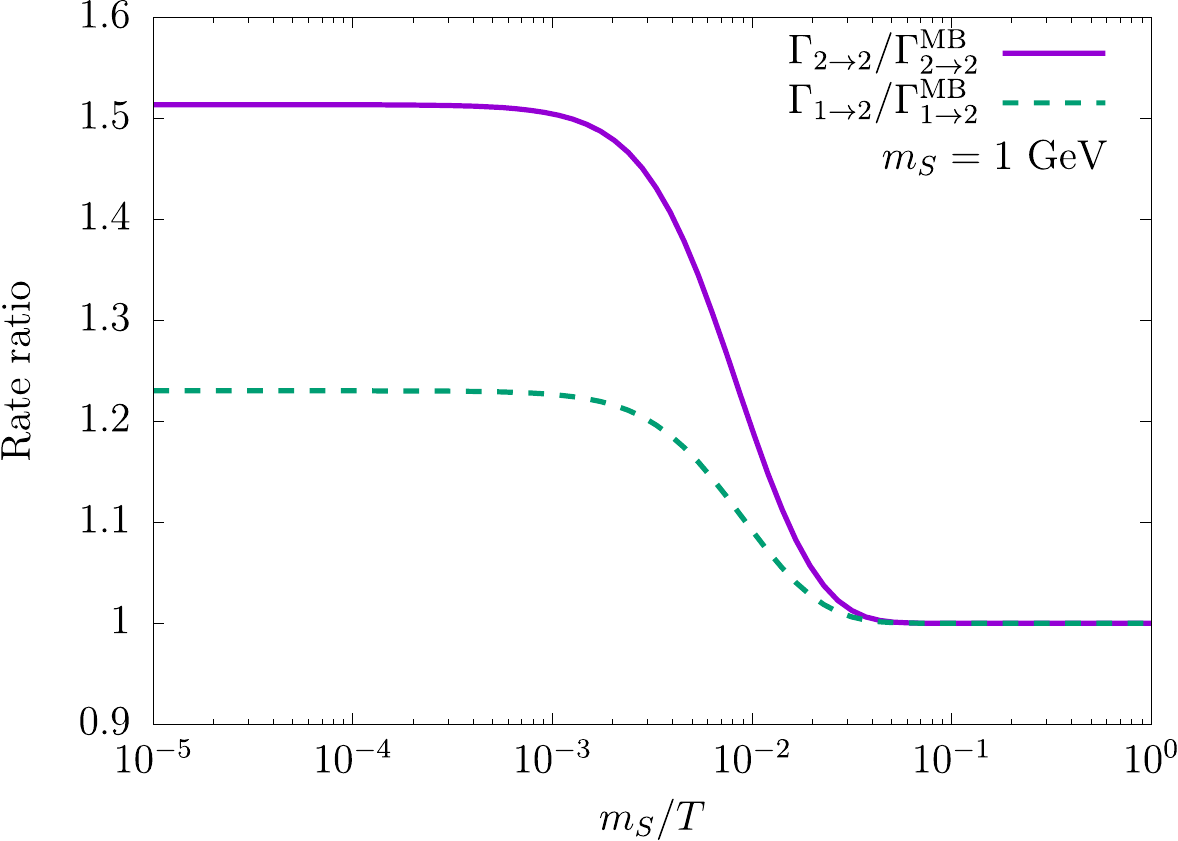}
 \includegraphics[scale=0.63]{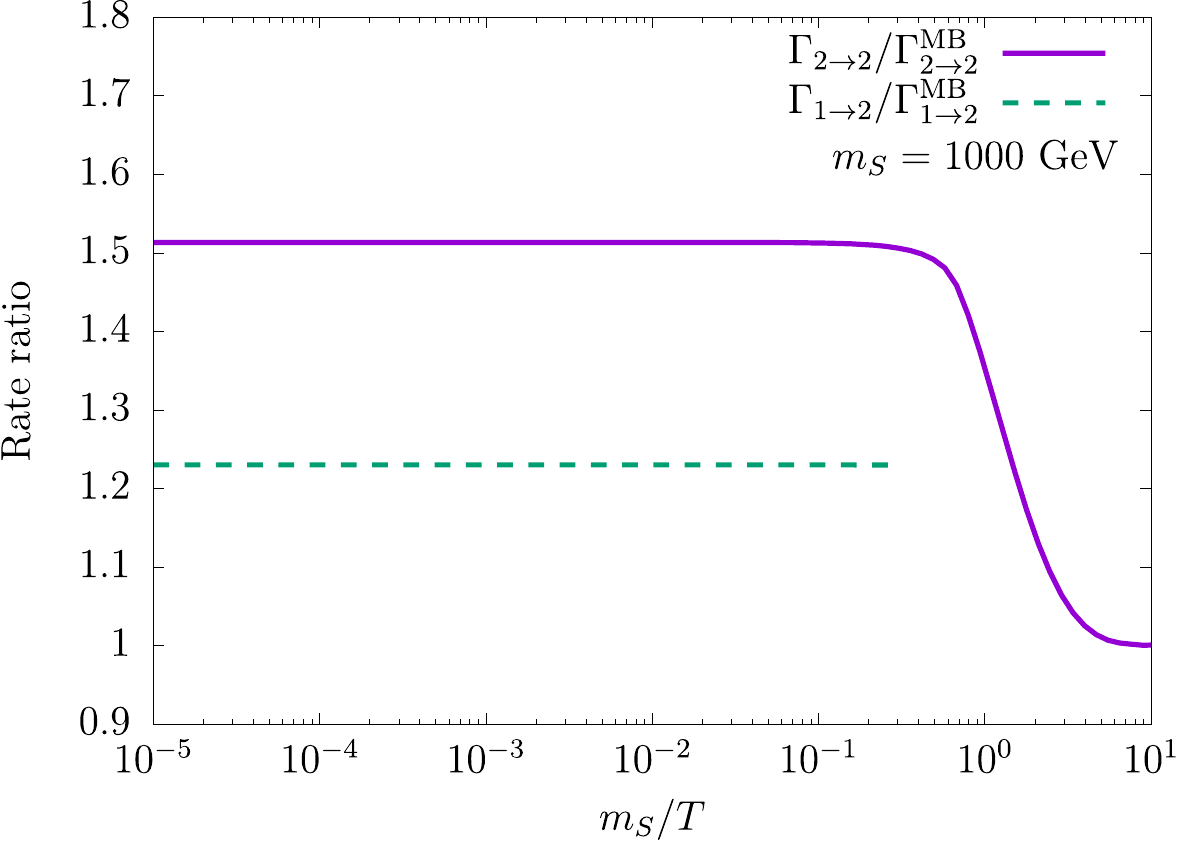}
\end{center}
\caption{   Comparison of the Bose--Einstein and  Maxwell--Boltzmann reaction rates in the Higgs thermal bath. \label{rates}}
\end{figure}

The computed reaction  rates are to be used in the Boltzmann equation in order to determine the DM density evolution.  Compared to their Maxwell--Boltzmann counterparts,
these rates are enhanced due to the Bose--Einstein distribution function peaking at low momenta. 
In general, the Bose--Einstein  rates can exceed the Maxwell--Boltzmann ones by orders of magnitude  \cite{Arcadi:2019oxh}, however 
the effect is sensitive to the thermal mass: for larger masses it is less pronounced.
In the case at hand, the Higgs field receives a large thermal correction due the gauge and top quark couplings.  The resulting enhancement is therefore modest as shown 
in Fig.\;\ref{rates}. It reaches 50\% for the annihilation mode and 20\% for the decay.

It is important to note that the inclusion of the thermal mass  regulates the high--$T$ behaviour of the rates which is equivalent to curing the infrared divergence as $m_{h,s} \rightarrow 0$.
Let us set $m_s=0$ and consider the limit $m_{h} \rightarrow 0$. In this case, we find that the   $2\rightarrow 2$ rate diverges as $\ln m_h$ which is unphysical. Including the thermal mass, we get  
\begin{equation}
\Gamma_{2\rightarrow 2} \propto T^4 \ln {T \over m_h} \rightarrow  c\; T^4 \;,
\end{equation}
which also represents the high--$T$ behaviour. Here $c$ is a constant depending on the couplings.
Therefore, the rate exhibits the expected scaling behaviour $ T^4$.

\section{The Boltzmann equation}

The evolution of the dark matter number density $n(t)$ is governed by the Boltzmann equation. In our case, it takes the form
\begin{equation}
 \dot{n} +3Hn = (4-3 \theta (T_{\rm EW} -T )) \times 2 \Gamma_{2\rightarrow 2}  + \theta (T_{\rm EW} -T) \times 2\Gamma_{1\rightarrow 2}\;,
 \label{Boltzmann-equation}
 \end{equation}
where the dot denotes a time derivative, $H$ is the Hubble rate; the factor of 2 is due to production of 2 DM particle in each reaction, and the $\theta$--functions take into account the change in Higgs degrees of freedom and the vertices at the electroweak phase transition. Here the inverse processes have been neglected given that the DM density is 
far below its value in equilibrium.

For our purposes, we may omit the contributions like $WW \rightarrow h \rightarrow ss$ (see e.g.\;\cite{Bernal:2018kcw}), which are only possible in the broken phase. We find that  if 
$m_h > 2m_s$, the DM yield is dominated by the decay in the broken phase and the annihilation processes can be neglected altogether. For $m_h < 2m_s$, the production takes place predominantly in the relativistic regime and the above mode is irrelevant. In conclusion, although the gauge boson contribution is similar in magnitude to $hh \rightarrow ss$ in  the broken phase, its effect is inconsequential.

 \subsection{Analytic solution in the relativistic regime}
 
 Qualitative behaviour of $n(t)$ in the relativistic regime  at $T> T_{\rm EW}$ can be understood analytically.
  It is convenient to trade the time variable for $T$. 
Due to SM entropy conservation,
$T^3 a^3 =const$, where $a$ is the scale factor. This implies $\dot T = -HT$ and the 
Boltzmann equation can be written as
\begin{equation}
T {dn \over dT} -3n + 8 {\Gamma_{2\rightarrow 2} \over H}=0 \;.
\label{Be}
\end{equation}
Here the Hubble rate is given by
\begin{equation}
 H= \sqrt{\pi^2 g_* \over 90}~ {T_{\rm }^2\over M_{\rm Pl}} \;,
 \end{equation}
with $g_*$
being the  number of  SM degrees of freedom and $M_{\rm Pl}$ being the reduced Planck mass. The last term in Eq.\;\ref{Be} scales as $T^2$ at high temperature, therefore it is convenient to define
a constant 
\begin{equation}
\kappa \equiv   8 {\Gamma_{2\rightarrow 2} \over H T^2}  \;.
\end{equation}
 Imposing the boundary condition that the DM abundance vanish at some initial temperature $T_0$, the solution to the Boltzmann equation reads
 \begin{equation}
n= \kappa  \;T^2 \left(  1- {T\over T_0} \right)\;.  
\label{n}
\end{equation}

We thus see that the total number of produced DM quanta  is proportional to $1/T$ at temperatures substantially below $T_0$. The result is conveniently expressed in terms of 
the ratio of the DM number density to the SM entropy density,
\begin{equation}
Y = {n\over s_{\rm SM}} ~~~,~~~ s_{\rm SM}={2 \pi^2  g_{*s}  \over 45}  \; {T^3}  \;, 
\end{equation}
 with $g_{*s}$ being the number of degrees of freedom contributing to the entropy. The correct DM relic abundance corresponds to
 \begin{equation}
  Y(\infty)=4.4 \times 10^{-10}~ \left(  {{\rm GeV} \over m_s } \right) \;. 
  \label{Yinfty}
  \end{equation}
 The total DM yield from (\ref{n}) depends on the temperature  $T_*$ at which relativistic production stops. 
 For $m_s > m_h$, we find that $T_* \simeq m_s$ and subsequent non--relativistic production is negligible. In that case, 
 Eq.\;\ref{Yinfty} requires  
 \begin{equation}
 \lambda_{hs} \simeq 2.2 \times 10^{-11}
\label{bound1}
\end{equation} 
 independently of $m_s$. Here we have used $g_* \simeq 107$ and also assumed  that the reheating temperature has been  high enough, $T \gg m_s$.
 
\begin{figure}[h!]
\begin{center}
 \includegraphics[scale=0.63]{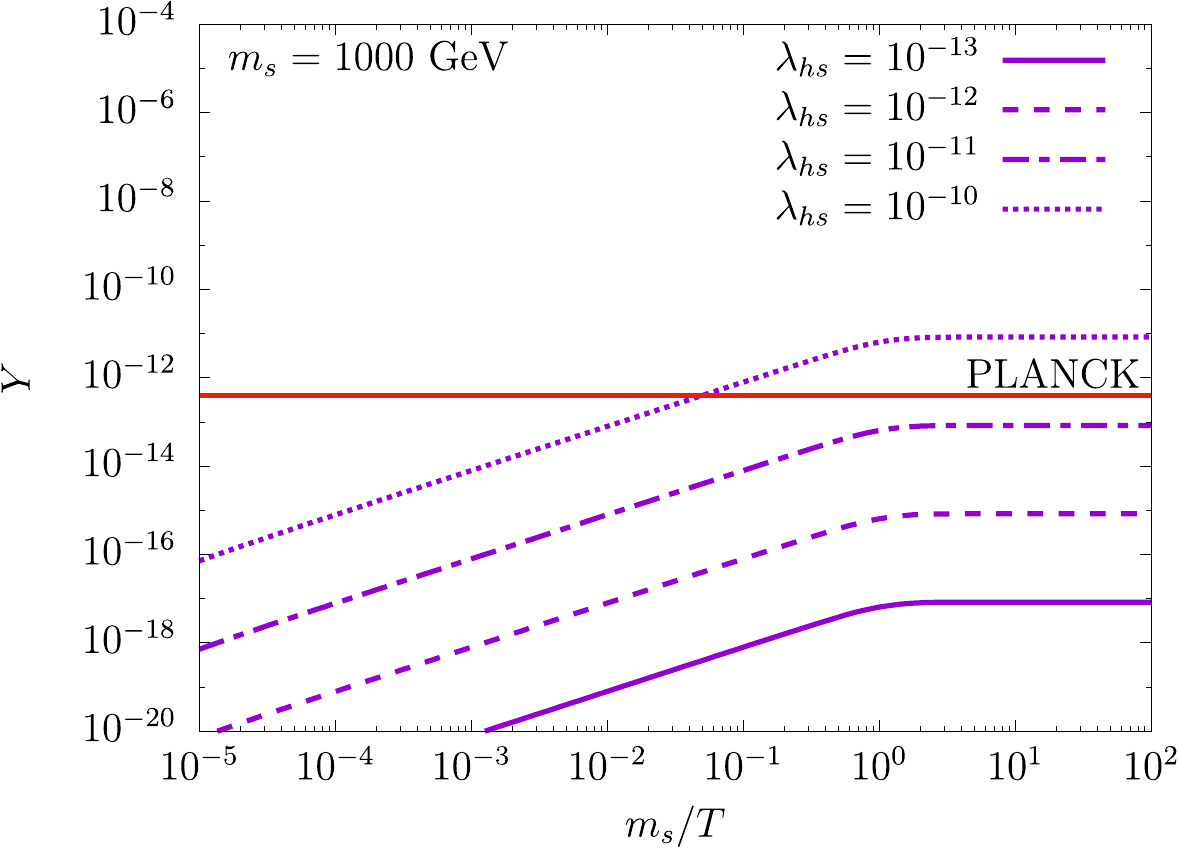}
 \includegraphics[scale=0.63]{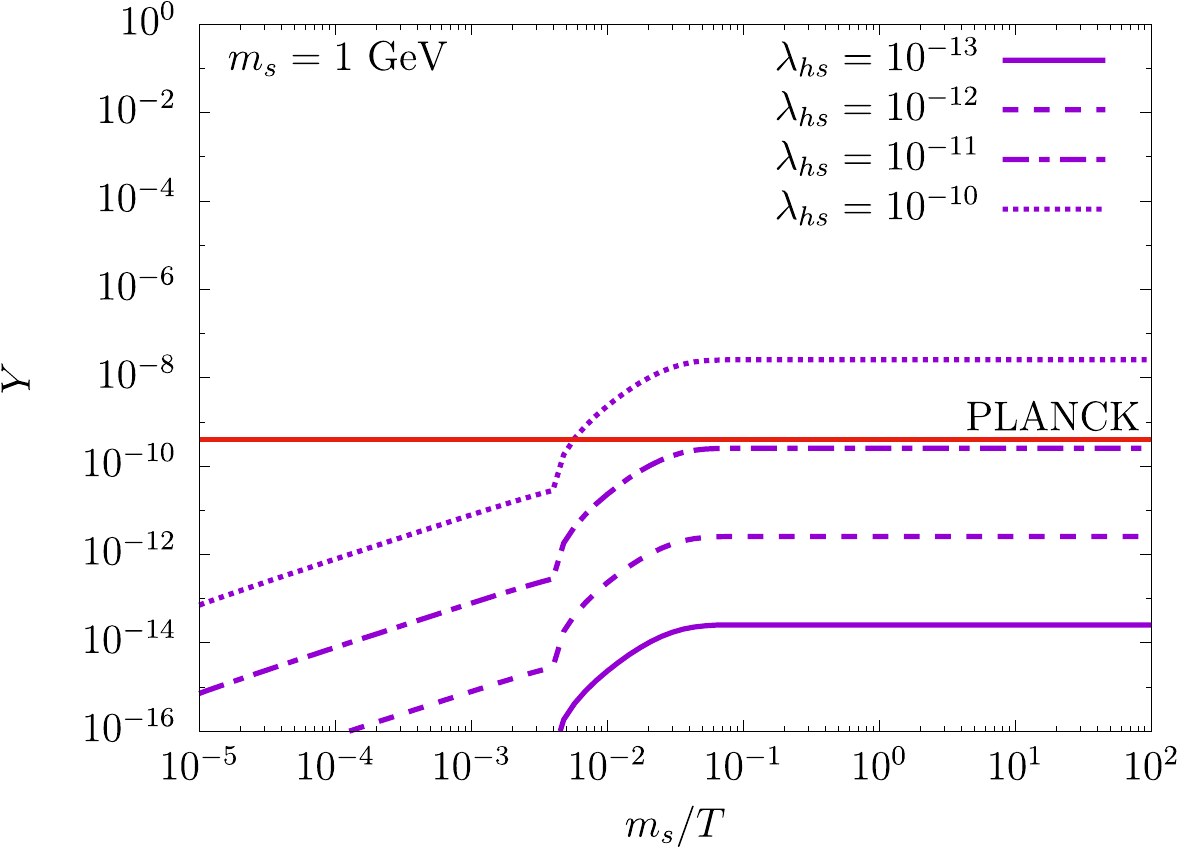}
\end{center}
\caption{   Numerical solution to the Boltzmann equation for different Higgs portal couplings.
The observed DM relic abundance is given by the red line. {\it Left:} 
DM production is dominated by Higgs annihilation  $h_i h_i\rightarrow ss$. 
{\it Right:} DM production is dominated by Higgs decay  $h \rightarrow ss$.
\label{Y(T)}}
\end{figure}
 
 Our numerical solution to the full Boltzmann equation (\ref{Boltzmann-equation}) supports this result. The evolution of $Y(T)$ 
 for representative values of $m_s$  is shown in Fig.\;\ref{Y(T)}. It grows as $1/T$ at high temperature and freezes--in at $T_* \sim m_s$ if $m_s> m_h$.
 For $m_s \ll m_h$, we find that the required coupling is 
 \begin{equation}
 \lambda_{hs} \simeq 1.2 \times 10^{-11}\; \sqrt{ {\rm GeV} \over m_s} \;.
 \label{bound2}
\end{equation} 
 In this case, the annihilation mode makes a negligible contribution. The decay mode activates at $T \lesssim T_{\rm EW}$ and dominates the DM production. At $T \lesssim 20$ GeV,
 the Higgs number density becomes too small and the process terminates.
 
 The numerical values of the necessary couplings (\ref{bound1}),(\ref{bound2}) are close to those obtained in \cite{Yaguna:2011qn}. The correction factors derived in our work tend to compensate each other leaving the original estimate almost intact.

 \subsection{Non--thermalization constraint}

An essential ingredient in our considerations is the assumption of non--thermalization of dark matter. Since thermal equilibrium implies detailed balance, 
we must require
\begin{equation}
\Gamma (hh \rightarrow ss) \gg \Gamma (ss \rightarrow hh) \;,
\end{equation}
which is satisfied for
\begin{equation}
n(T) \ll n_{\rm eq}(T) \;,
\end{equation}
where $n_{\rm eq}(T) $ is the equilibrium density at the temperature of the SM thermal bath.
Our solution  (\ref{n}) for $n(T)$ then implies the constraint
\begin{equation}
\lambda_{hs} \ll 10^{-7}   \; \sqrt{m_s\over {\rm GeV}} \;,
\end{equation}
 where we have taken  $T \sim m_s$ as the lowest temperature at which relativistic production can take place. The couplings in the range of interest satisfy this constraint.

The thermalization bound on $\lambda_s$ has recently been obtained in \cite{Arcadi:2019oxh}. The calculation is quite complicated since it involves a $2\rightarrow 4$ process and
depends on the DM number density,
while for our purposes it is sufficient to take  a conservative bound $\lambda_s < 10^{-5}$.

\section{Conclusion}

The aim of this work is to study  the freeze--in production of dark matter in the relativistic regime, focussing on the simplest Higgs portal model.
To this end, we have derived the relativistic reaction rates with the Bose--Einstein distribution function and solved analytically as well as numerically  the corresponding Boltzmann equation.
Obtaining a physically meaningful result requires inclusion of the Higgs thermal mass correction.

We have also made several  other improvements over previous analyses. In particular, we take into account the electroweak phase transition. This implies, for example,
that above the critical temperature the trilinear Higgs--DM vertices do not exist, which forbids Higgs decay and gauge boson annihilation into dark matter. The relevant DM production
mode is instead annihilation of 4 massive Higgs degrees of freedom through the Higgs portal  vertex.  The required couplings
leading to the correct relic abundance are
 given by Eqs.\;\ref{bound1},\ref{bound2}.

The overall numerical effect of our improvements happens to be  modest compared to a straightforward (albeit unjustified) extrapolation of the non--relativistic result to high temperatures.
Nevertheless, our systematic approach results in  a more {\it reliable} evaluation of the dark matter yield and the relevant constraints on parameter space.

{}

\end{document}